\documentstyle[aps,psfig]{revtex}
\begin{document}
\title{Quantum mutual entropy for two-level ion in a q-analog trap}
\author{S. Shelly Sharma}
\address{Depto. de Fisica, Universidade Estadual de Londrina, Londrina, 86040-370,\\
PR, Brazil}
\author{N. K. Sharma}
\address{Depto. de Matematica, Universidade Estadual de Londrina, Londrina, 86040-370,%
\\
PR, Brazil}
\maketitle

\begin{abstract}
Quantum mutual entropy is used as a measure of information content of ionic
state due to ion-laser interaction in a q-analog trap. The initial state of
the system is a Schrodinger cat state. It is found that the partial mutual
entropy is a good measure of the entanglement and purity of the ionic state
at $t>0$.
\end{abstract}


New insight into quantum mechanical systems has been obtained by applying
information theory technics \cite
{Ohya93,Benn96,Vedr97,Pano97,Bela98,Ohya89,Umeg62}. In particular different
probability distributions may be compared by using quantum relative entropy%
\cite{Vedr97} and the degree of entanglement measured by quantum mutual
entropy\cite{Bela98}. In this article we examine the entropy, quantum
relative entropy and quantum mutual entropy for trapped ion- laser system,
as a measure of entanglement and quantum information propagation. For an
excellent review on quantum relative entropy see Ref. \cite{Vedr01}.

We consider an ion interacting with a single mode laser in a q-analog
harmonic oscillator trap with the system Hamiltonian given by 
\begin{equation}
H_{q}=\frac{1}{2}\hbar \omega (AA^{\dagger }+A^{\dagger }A)+\frac{1}{2}\hbar
\triangle \sigma _{z}+\frac{1}{2}\hbar \Omega (F_{q}\sigma
^{+}+F_{q}^{\dagger }\sigma ^{-})\text{ ,}  \label{eq1}
\end{equation}
where $A^{\dagger }$ and $A$ are the creation and annihilation operators for
the trap quanta satisfying the commutation relations,

\begin{equation}
AA^{\dagger }-qA^{\dagger }A=q^{-N}\hspace{0.2in};\hspace{0.2in}NA^{\dagger
}-A^{\dagger }N=A^{\dagger }\hspace{0.2in};\hspace{0.2in}NA-AN=-A,
\label{eq2}
\end{equation}
$q$ the trap deformation parameter ($q=\exp (\tau )$ with $\tau $ $%
%TCIMACRO{\func{real}}
%BeginExpansion
\mathop{\rm real}%
%EndExpansion
$)., $\Delta =\omega _{a}-\omega _{l}$ the detuning parameter, and $\Omega $
the Rabi frequency. The pseudo-spin operators of the two-level ion, $\sigma
^{\pm }$ and $\sigma _{z}$ operate on state vectors $\left| g\right\rangle $
and $\left| e\right\rangle $ where indices $g$ and $e$ stand for the ground
and excited states of the two level ion. The operator $F_{q}=e^{\left( \frac{%
-\left| \epsilon \right| ^{2}}{2}\right) }e^{i\epsilon A^{\dagger
}}e^{i\epsilon A}$ , and $N$ is the number operator for trap quanta, .

Treating internal and motional degrees of freedom as separate systems, the
entropies are defined through the corresponding reduced density operators by 
\begin{equation}
S_{ion(c.m.)}=-tr_{ion(c.m.)}\left\{ \rho _{ion(c.m.)}\ln (\rho
_{ion(c.m.)})\right\}  \label{eq3}
\end{equation}
where the reduced density operators are 
\begin{equation}
\rho _{ion(c.m.)}=tr_{c.m.(ion)}\left\{ \rho \right\} .  \label{eq4}
\end{equation}

We consider the initial state of the trapped ion to be a pure state with 
\begin{equation}
\rho _{0}^{(1)}=\left| g\right\rangle \left\langle g\right| \otimes \left|
\beta \right\rangle _{q}\ _{q}\left\langle \beta \right| \text{ , or }\rho
_{0}^{(2)}=\left| e\right\rangle \left\langle e\right| \otimes \left| -\beta
\right\rangle _{q}\ _{q}\left\langle -\beta \right| \text{ ,}  \label{eq5}
\end{equation}
where the motional state of the ionic center of mass is a $q-$ analog
coherent state 
\begin{equation}
\left| \beta \right\rangle _{q}=\frac{1}{\sqrt{\exp _{q}\left( \left| \beta
\right| ^{2}\right) }}\sum\limits_{n=0}^{\infty }\frac{\beta ^{n}}{\sqrt{%
\left[ n\right] _{q}!}}\left| n\right\rangle _{q}\quad .  \label{eq6}
\end{equation}
A pure state at $t=0$ evolves into an entangled state for $t>0$ with the
Hamiltonian $H_{q}$ governing the dynamics of the system through 
\begin{equation}
H_{q}\Psi ^{(i)}(t)=i\hbar \frac{d}{dt}\Psi ^{(i)}(t)\text{ }  \label{eq7}
\end{equation}
($i=1$ for ionic state $\left| g\right\rangle $ and for ionic state $\left|
e\right\rangle $).

Working in the space spanned by basis vectors $\{\left| u_{n}\right\rangle
\}\equiv \left| g,m\right\rangle ,\left| e,m\right\rangle ,m=0,1,.....\infty
,$ the state of the system may be written as 
\begin{equation}
\Psi ^{(i)}(t)=\sum\limits_{m}g_{m}^{(i)}(t)\left| g,m\right\rangle
+\sum\limits_{m}e_{m}^{(i)}(t)\left| e,m\right\rangle \quad .  \label{eq8}
\end{equation}
We next define the populations and coherences for the system prepared in
initial state $\left| i,\beta \right\rangle _{q}$, with the populations
defined as 
\begin{equation}
P_{g}^{(i)}(t)=\left\langle g\left| tr_{ion}\rho _{{}}^{(i)}(t)\right|
g\right\rangle =\sum_{n}\left| g_{n}^{(i)}(t)\right| ^{2}  \label{eq9}
\end{equation}
and 
\begin{equation}
P_{e}^{(i)}(t)=\left\langle e\left| tr_{ion}\rho _{{}}^{(i)}(t)\right|
e\right\rangle =\sum_{n}\left| e_{n}^{(i)}(t)\right| ^{2}.  \label{eq10}
\end{equation}
The coherences for this initial states are given by 
\begin{equation}
C_{ge}^{(i)}(t)=\left\langle g\left| tr_{ion}\rho _{{}}^{(i)}(t)\right|
e\right\rangle =\sum_{n}\left( g_{n}^{(i)}(t)\right)
^{*}e_{n}^{(i)}(t)\qquad .  \label{eq11}
\end{equation}
and 
\begin{equation}
C_{eg}^{(i)}(t)=\left\langle e\left| tr_{ion}\rho _{{}}^{(i)}(t)\right|
g\right\rangle =\left( C_{ge}^{(i)}(t)\right) ^{*}.  \label{eq12}
\end{equation}
A straightforward calculation shows that 
\begin{eqnarray}
S_{ion}^{(i)} &=&-\left[ P_{g}^{(i)}(t)\log P_{g}^{(i)}(t)+P_{e}^{(i)}\log
P_{e}^{(i)}(t)\right.  \nonumber \\
&&\left. +2*%
%TCIMACRO{\func{Re} }
%BeginExpansion
\mathop{\rm Re}%
%EndExpansion
C_{ge}^{(i)}(t)\log \left( C_{ge}^{(i)}(t)\right) ^{*}\right] \text{ .}
\label{eq13}
\end{eqnarray}
As at $t=0,$ ionic state and center of mass motional state are pure states,
the total entropy of the system is zero. At $t>0$ the entropies of the
center of mass motion and the ionic subsystems are precisely equal. The
entropy $S_{ion}$ is a good a measure of entanglement of the system and has
more information content than the population inversion which is an
experimentally measurable property of the system.

Next we consider trapped ion prepared in the initial state $\Psi (t=0)=\frac{%
\left| g,\beta \right\rangle _{q}+\ \left| e,-\beta \right\rangle _{q}}{%
\sqrt{2}}$ , which is a Macroscopic superposition of states. This is not a
pure state but in the phase space the components $\left| g,\beta
\right\rangle _{q}$ and $\left| e,-\beta \right\rangle _{q}$ are well
separated and distinguishable. The density operator for the internal states
of the ion at $t=0$ is 
\begin{equation}
\rho _{ion}(t=0)=\frac{1}{2}\left( \left| g\right\rangle \left\langle
g\right| +\left| e\right\rangle \left\langle e\right| \right) .  \label{eq14}
\end{equation}
The state of the system at a time $t$ given 
\begin{equation}
\Psi (t)=\sum\limits_{m}g_{m}(t)\left| g,m\right\rangle
+\sum\limits_{m}e_{m}(t)\left| e,m\right\rangle  \label{eq15}
\end{equation}
is a solution of the time dependent Schrodinger equation 
\begin{equation}
H_{q}\Psi (t)=i\hbar \frac{d}{dt}\Psi (t).  \label{eq16}
\end{equation}
Defining the probabilities of finding the ion in the ground state and the
excited state (the populations) to be respectively 
\begin{equation}
P_{g}(t)=\left\langle g\left| tr_{ion}\rho (t)\right| g\right\rangle
=\sum_{n}\left| g_{n}(t)\right| ^{2}  \label{eq17}
\end{equation}
and 
\begin{equation}
P_{e}(t)=\left\langle e\left| tr_{ion}\rho (t)\right| e\right\rangle
=\sum_{n}\left| e_{n}(t)\right| ^{2}.  \label{eq18}
\end{equation}
and the non-diagonal matrix element of the operator $tr_{ion}\rho (t)$
(coherence), 
\begin{equation}
C_{ge}=\left\langle g\left| tr_{ion}\rho (t)\right| e\right\rangle
=\sum_{n}g_{n}^{*}(t)e_{n}(t)\qquad .  \label{eq19}
\end{equation}
We find that the entropy 
\begin{eqnarray}
S_{ion} &=&-\left[ P_{g}(t)\log P_{g}^{{}}(t)+P_{e}^{{}}\log
P_{e}^{{}}(t)\right.  \nonumber \\
&&\left. +2*%
%TCIMACRO{\func{Re} }
%BeginExpansion
\mathop{\rm Re}%
%EndExpansion
C_{ge}^{{}}(t)\log \left( C_{ge}^{{}}(t)\right) ^{*}\right] \text{ .}
\label{20}
\end{eqnarray}

We plot in figure I(a) $Sg=S_{ion}^{(1)}$ for $\tau =0.0$ ( Harmonic
oscillator trap) , figure I(b) $Scat=S_{ion}$ for $\tau =0.0$ , and figure
I(c) $Scat=S_{ion}$ for $\tau =0.004$ ( $q-$deformed harmonic oscillator
trap) as a function of rescaled time parameter $t(\frac{\Omega t}{2\pi })$
for $\beta =4.$ The parameter values used in the numerical calculation are $%
\overline{\omega }=\frac{\omega }{\Omega }=50,$ $\overline{\epsilon }=\frac{%
\epsilon }{\Omega }=0.05,$ and $\overline{\Delta }=\frac{\Delta }{\Omega }%
=-50$. The maximum value of $n$ in Eq. (\ref{eq6}) is restricted to $n=32$.
For the case (a) where initial state of the ion is a pure state with $Sg=0$,
the value of $Sg$ increases with $t$ indicating entanglement. After an
interval of time characteristic of the system $Sg$ is found to go to zero
again indicating a decoupling of the internal and motional degrees of
freedom of the ion. In case (b) $Scat$ follows more or less the population
inversion. For the case (c) no conclusive information can be obtained about
the state of the system at hand.

In order to measure the quantum interference effects due to the initial
state superposition of component states $\left| g,\beta \right\rangle _{q}$
and $\left| e,-\beta \right\rangle _{q}$ we use the notion of quantum mutual
entropy. Furuichi and co-workers\cite{furu99} have applied the concept of
quantum mutual entropy to dynamical change of state of the atom in
Jaynes-Cummings model(JCM)\cite{Jayn63}. Quantum relative entropy \cite
{Umeg62,Ohya93} between the two states $\sigma $ and $\rho $ defined as 
\begin{equation}
S(\sigma ,\rho )=Tr\ \sigma (\ln \sigma -\ln \rho )\text{ ,}  \label{eq21}
\end{equation}
is a measure of difficulty of distinguishing between the states $\sigma $
and $\rho $. The correlation between the initial state $\rho _{ion}(t=0)$
having component states $\left| i\right\rangle \left\langle i\right| $ ($%
i=1,2$) and the state $\rho _{ion}(t)$ at $t>0$ can be measured by quantum
relative entropies 
\begin{equation}
S(\sigma _{1}^{(i)},\sigma _{2}^{{}})=tr\sigma _{1}^{(i)}(\ln \sigma
_{1}^{(i)}-\ln \sigma _{2})\text{ ;\ }(i=1,2)\text{ , where the states }
\label{22}
\end{equation}
\begin{equation}
\sigma _{1}^{(i)}=\frac{1}{2}\left| i\right\rangle \left\langle i\right|
\otimes \rho _{ion}^{(i)}(t)\text{ and }\sigma _{2}=\rho _{ion}(t=0)\otimes
\rho _{ion}^{{}}(t).  \label{23}
\end{equation}
The quantum mutual entropy \ 
\begin{eqnarray}
I\left( \rho _{ion}(t=0),\rho _{ion}^{{}}(t)\right)  &=&\sum_{i=1,2}\frac{1}{%
2}S(\sigma _{1}^{(i)},\sigma _{2}^{{}})  \nonumber \\
&=&\sum_{i=1,2}\frac{1}{2}\sum_{k=1,2}\left[ \left\langle g\right| \sigma
_{1}^{(i)}\left| k\right\rangle \left( \ln \frac{\left\langle k\right|
\sigma _{1}^{(i)}\left| g\right\rangle }{\left\langle k\right| \sigma
_{2}\left| g\right\rangle }\right) \right.   \nonumber \\
&&\left. +\left\langle e\right| \sigma _{1}^{(i)}\left| k\right\rangle
\left( \ln \frac{\left\langle k\right| \sigma _{1}^{(i)}\left|
e\right\rangle }{\left\langle k\right| \sigma _{2}\left| e\right\rangle }%
\right) \right] \text{ .}  \label{24}
\end{eqnarray}
measures the degree of entanglement of the state $\rho _{ion}^{{}}(t)$. We
can verify that in terms of the populations and the coherences defined above
the relative quantum entropies for the system starting in initial state
given by Eq. (\ref{eq14}) are 
\begin{eqnarray}
S(\sigma _{1}^{(i)},\sigma _{2}^{{}}) &=&\left[ P_{g}^{(i)}(t)\log \left( 
\frac{P_{g}^{(i)}(t)}{P_{g}^{{}}(t)}\right) +P_{e}^{(i)}\log \left( \frac{%
P_{e}^{(i)}(t)}{P_{e}^{{}}(t)}\right) \right.   \nonumber \\
&&\left. +2*%
%TCIMACRO{\func{Re}}
%BeginExpansion
\mathop{\rm Re}%
%EndExpansion
\left[ C_{ge}^{(i)}(t)\log \left( \frac{C_{ge}^{(i)}(t)}{C_{ge}^{{}}(t)}%
\right) ^{*}\right] \right] \text{ .}  \label{25}
\end{eqnarray}
Substituting Eq. (\ref{25}) in the definition of quantum mutual entropy for
the system, we have 
\begin{eqnarray}
I\left( \rho _{ion}(t=0),\rho _{ion}^{{}}(t)\right)  &=&\sum_{i=1,2}\frac{1}{%
2}\left[ P_{g}^{(i)}(t)\log \left( \frac{P_{g}^{(i)}(t)}{P_{g}^{{}}(t)}%
\right) +P_{e}^{(i)}\log \left( \frac{P_{e}^{(i)}(t)}{P_{e}^{{}}(t)}\right)
\right.   \nonumber \\
&&\left. +2*%
%TCIMACRO{\func{Re}}
%BeginExpansion
\mathop{\rm Re}%
%EndExpansion
\left[ C_{ge}^{(i)}(t)\log \left( \frac{C_{ge}^{(i)}(t)}{C_{ge}^{{}}(t)}%
\right) ^{*}\right] \right] \qquad .  \label{26}
\end{eqnarray}
We can split $I\left( \rho _{ion}(t=0),\rho _{ion}^{{}}(t)\right) $ in to
two distinct parts, one depending on populations and the other on off
diagonal coherences that is 
\begin{equation}
I\left( \rho _{ion}(t=0),\rho _{ion}^{{}}(t)\right) =S(P)+S(C)  \label{27}
\end{equation}
where 
\begin{equation}
S(P)=\sum_{i=1,2}\lambda (i)\left[ P_{g}^{(i)}(t)\log \left( \frac{%
P_{g}^{(i)}(t)}{P_{g}^{{}}(t)}\right) +P_{e}^{(i)}(t)\log \left( \frac{%
P_{e}^{(i)}(t)}{P_{e}^{{}}(t)}\right) \right] \text{ .}  \label{28}
\end{equation}
We define population inversion, the difference between the probabilities of
finding the system in the ground state and the excited state, as 
\[
Inv=P_{g}(t)-P_{e}(t),
\]
which is the most commonly used measure of entanglement for ion in trap
system.

Figure II is a plot of quantum mutual entropy $I$ , partial quantum mutual
entropy $S(P)$ and population inversion $Inv$ as a function of rescaled time
parameter $t(\frac{\Omega t}{2\pi })$ for $\beta =4.$ The $q$-deformed
oscillator trap is characterized by deformation parameter $\tau =0.004$ ($%
q=\exp (\tau )$ with $\tau $ $%
%TCIMACRO{\func{real}}
%BeginExpansion
\mathop{\rm real}%
%EndExpansion
$). This particular system, studied in refs. \cite{Shel97,Shel01} develops
extremely well defined collapses and revivals\cite{Cumm65,Shor93} of
population inversion as is evident from figure II(a). The quantum mutual
entropy $I$ measures the system correlations at collapse, revival peaks and
during the transition period when the revival is building up or decaying. As
expected the transition periods correspond to highly correlated system. The
partial quantum mutual entropy $S(P)$ containing no contributions from
coherences shows a very regular pattern with peaks characterizing each
collapse and revival. The onset of \ each collapse and revival is
characterized by $S(P)=0$ and fluctuating $I$ $\ $value. The $S(p)$ peaks
represent the $t$values for which the information content of the ionic
system is the maximum. With time quantum mutual entropy $I$ \ as well as $%
S(P)$ are seen to decrease indicating a dissipative irreversible change in
the ionic state. In ref. \cite{furu99}, where quantum mutual entropy
dynamics for Jaynes-Cummings model has been reported the coherences turn out
to be zero ( $S(P)=I$). As such our results for partial mutual quantum
entropy $S(P)$ are similar to the results for total quantum mutual entropy $I
$ of ref. \cite{furu99}. From the location of $S(P)$ peaks\ the successive $%
Inv$ collapses occur at $t=67.8,201.2,$ and $336.8$ while the revival peaks
are seen at $t=133.2$ and $266.6$ .

{\Large Acknowledgments}

S. S. S. and N. K. S. would like to acknowledge financial support from
Universidade Estadual de Londrina, Brazil.

\begin{figure}
\newpage
\psfig{figure=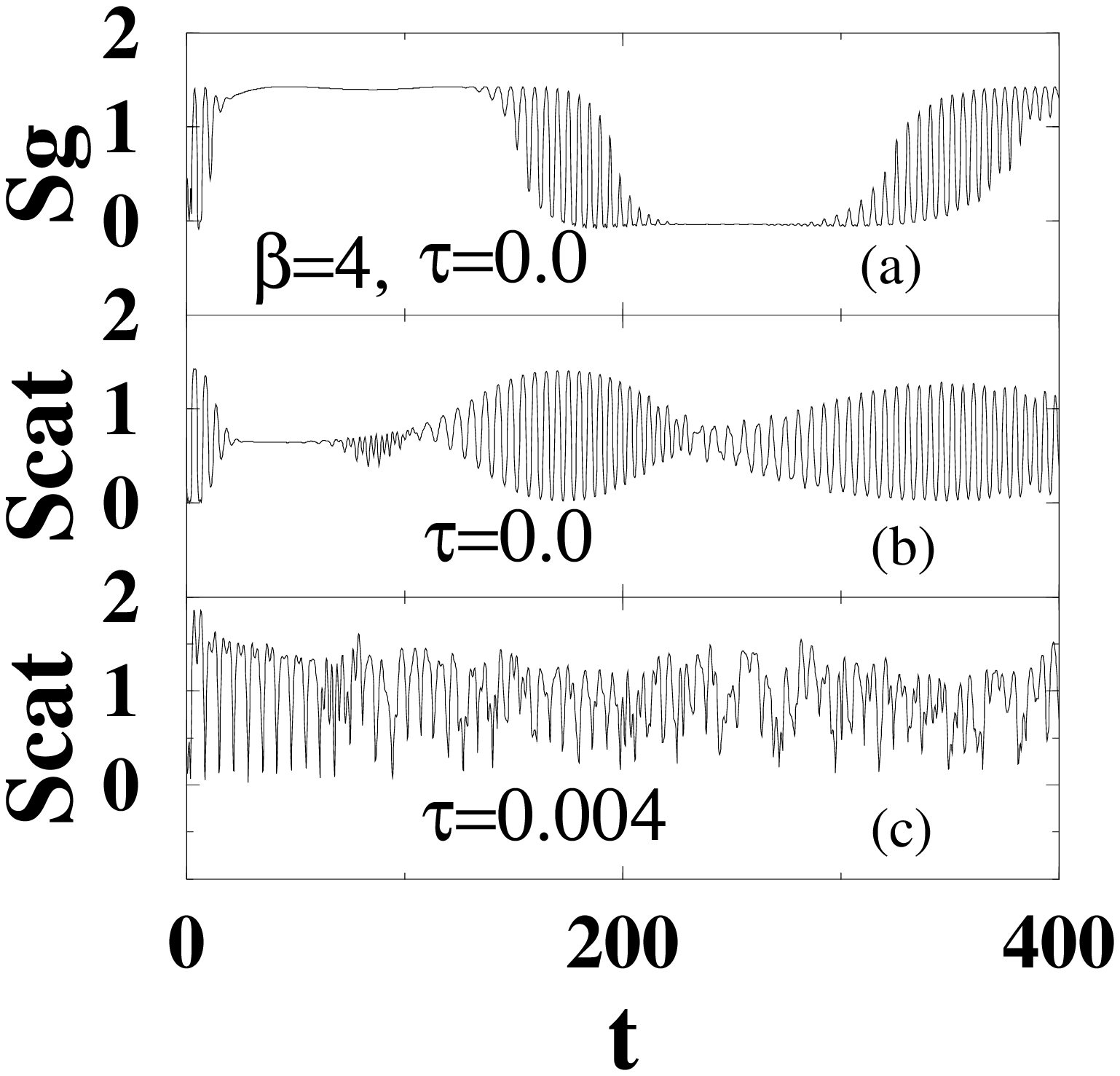,width=6in,height=7in}
\caption{(a) $Sg=S_{ion}^{(1)}$ for $\tau =0.0$( Harmonic oscillator trap),
(b) $Scat=S_{ion}$ for $\tau =0.0$ , and (c) $Scat=S_{ion}$ for $\tau =0.004$
( $q-$deformed harmonic oscillator trap) as a function of rescaled time
parameter $t(\frac{\Omega t}{2\pi })$ for $\beta =4.$}
\label{fig1}
\end{figure}
\newpage 
\begin{figure}
\psfig{figure=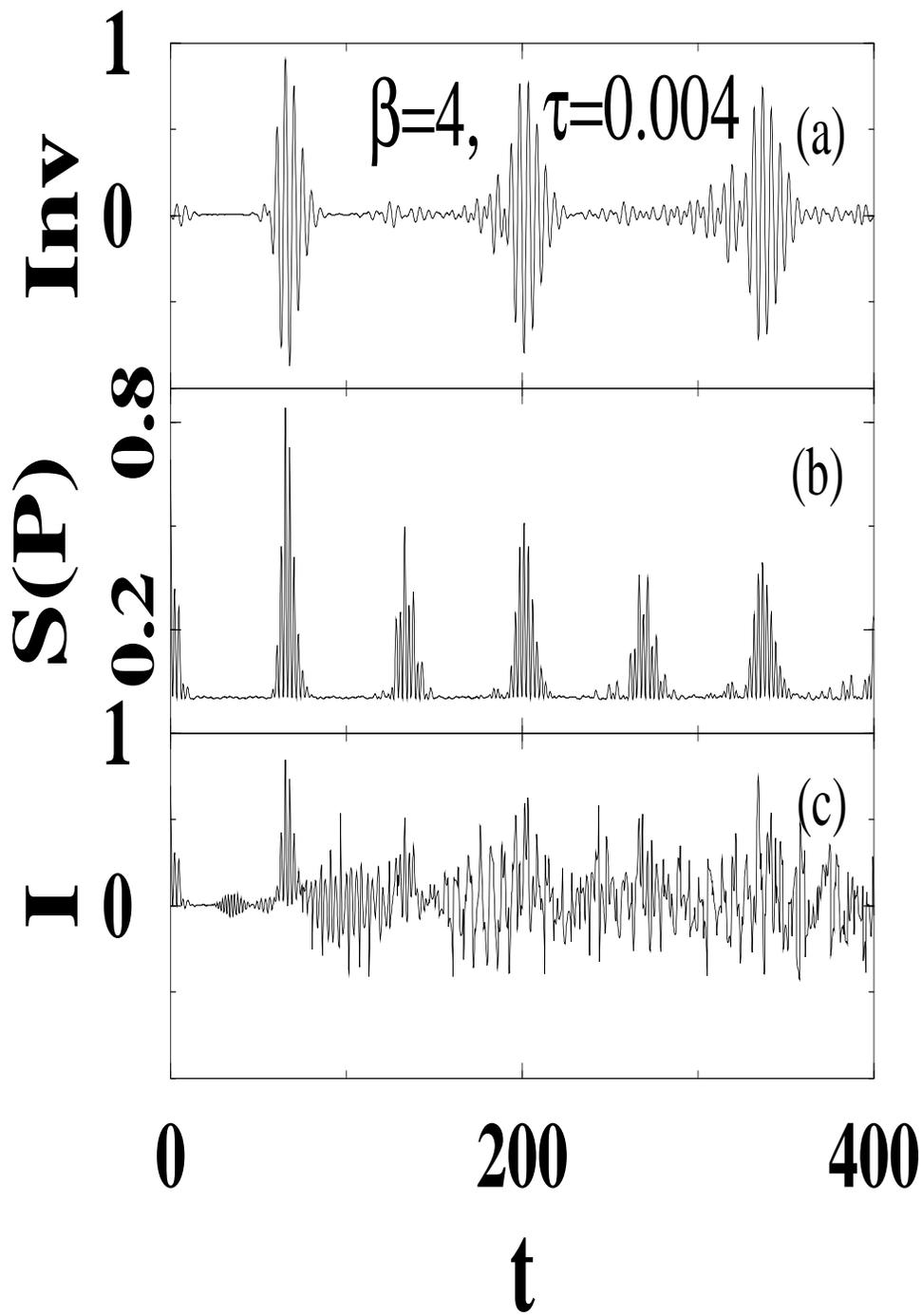,width=6in,height=7in}
\caption{Quantum mutual entropy $I$ , partial quantum mutual entropy $S(P)$
and population inversion $Inv$ as a function of rescaled time parameter $t(%
\frac{\Omega t}{2\pi })$ for $\beta =4$ and $\tau =0.004$.}
\label{fig2}
\end{figure}

\end{document}